\documentclass[fleqn]{article}
\usepackage{longtable}
\usepackage{graphicx}

\begin {document}
\begin{flushleft}
{\LARGE
{\bf Radiative rates and electron impact excitation rates for transitions in He II}
}\\

\vspace{1.5 cm}

{\bf {K. M. Aggarwal$^1$, A. Igarashi$^2$, F. P. Keenan$^1$, and S. Nakazaki$^2$}}\\ 

\vspace*{1.0cm}

\vspace*{0.5 cm} 
$^1$Astrophysics Research Centre, School of Mathematics and Physics, Queen's University Belfast, \\Belfast BT7 1NN, Northern Ireland, UK\\K.Aggarwal@qub.ac.uk \\
$^2$Department of Applied Physics, Faculty of Engineering, University of Miyazaki, Miyazaki 889-2192, Japan\\
e-mail: K.Aggarwal@qub.ac.uk \\

\vspace*{0.20cm}

Received  26 January 2017\\  Accepted for publication 20 April  2017 \\ Published xx Month  2017 \\ 

\vspace*{1.5cm}


\vspace*{1.0 cm}

\hrule

\vspace{0.5 cm}

\end{flushleft}

\clearpage


\begin{abstract}

We report  calculations of energy levels, radiative rates, collision strengths, and effective collision strengths for 
transitions among the lowest 25 levels of the $n \le$ 5 configurations of He~II.
The general-purpose relativistic atomic structure package ({\sc grasp}) and Dirac atomic R-matrix code ({\sc darc}) are adopted for the calculations. 
Radiative rates, oscillator strengths, and line strengths are reported for all electric dipole (E1), magnetic dipole (M1), electric quadrupole (E2), and magnetic 
quadrupole (M2) transitions among the 25 levels. Furthermore, collision strengths and effective  collision strengths are listed for all 300 transitions among the above 25
levels over a wide energy (temperature) range up to 9 Ryd (10$^{5.4}$ K). Comparisons are made with earlier available results and the accuracy of the data is assessed.

\end{abstract}

\clearpage

\section{Introduction}

Helium is the second most abundant element in the Universe. Emission lines of He~II have been observed in the Sun and other  astrophysical sources, such as early-type
stars, gaseous nebulae, and active galaxies. Many of the observed emission lines from He~II are listed in the NIST ({\tt http://physics.nist.gov/PhysRefData}) and {\sc
chianti}  ({\tt http://www.chiantidatabase.org/}) databases. 

The analysis of emission or absorption lines from a plasma  provides information on its physical properties,  such as  temperature, density and chemical composition. However, such
an analysis requires information for a wide range of atomic parameters, including energy levels, radiative rates and excitation rate coefficients. Therefore, in this paper we
report calculations for transitions in H-like He~II.

Apart from energy levels, there is a paucity of measurements for the above atomic parameters for He~II, although some early results for cross sections are available by
Dolder and  Peart \cite{dp} for the 1s-2s transition. Therefore, theoretical results are required to reliably analyse plasma spectra.  A few calculations have been performed
in the past, the most notable being those of Aggarwal et al. \cite{he2a}, Kisielius et al. \cite{kbn96}, and Ballance et al. \cite{bbs03}. 

Aggarwal et al. \cite{he2a} performed non-relativistic calculations in $LS$ coupling for transitions among the $n \leq$ 5 states. They adopted the $R$-matrix program of
Berrington et al. \cite{rm1}, and resolved resonances in the threshold region  to include their contribution to the effective collision strengths, $\Upsilon$. However, since it
is the {\em fine-structure} transitions which are observed spectroscopically, their calculations were of limited application. This limitation was removed by Kisielius et al.
\cite{kbn96}, who performed fully relativistic calculations in $jj$ coupling. These authors  also resolved resonances in the threshold region, and employed the earlier version of the
{\em Dirac atomic R-matrix code} ({\sc darc}). However, their calculations suffer from a few limitations. Firstly, their results for $\Upsilon$ were restricted to transitions
among the $n \le$ 4 levels, whereas data for many transitions involving the $n$ = 5 levels are also required (see the NIST and/or {\sc chianti} databases). Secondly, they
did not report results for `elastic' (i.e. allowed with $\Delta n$ = 0) transitions.  Thirdly, the corresponding data for radiative rates were not provided. These are
required, along with the excitation rates, in any modelling application. Finally, and most importantly, their calculations for collision strengths ($\Omega$) were limited to 
energies below 7 Ryd, and for the effective collision strengths ($\Upsilon$) to temperatures below  10$^{4.3}$ K, whereas the temperature of maximum abundance in ionisation
equilibrium for He~II is 10$^{4.7}$ K -- see Bryans et al. \cite{pb}. Therefore, there is a clear need to extend the calculations of Kisielius et al.

Ballance et al. \cite{bbs03} also adopted the $R$-matrix approach and considered the same 15 $LS$ states (within $n \le$ 5) as  Aggarwal et al. \cite{he2a}. Furthermore,
they  resolved resonances in the thresholds region and considered a wide range of partial waves with angular momentum $L \le$ 60, comparable to the   $L \le$ 45 by Aggarwal
et al., and $J \le$ 40 by Kisielius et al. \cite{kbn96}. Additionally, in the expansion of the wavefunctions they included 24 {\em pseudostates} together with the 15 {\em
physical} states. As well as  performing their calculations in $LS$ coupling, they did not report any results {\em except} for values of $\Upsilon$, and even then for only five
transitions at four temperatures in the range 3.2 $\le$ log T$_e$ (K) $\le$ 4.3. However, data for $\Upsilon$ for transitions from the lowest two levels (1s $^2$S and 2s
$^2$S) to higher excited levels are available in the {\sc chianti} database. 

In conclusion, the most comprehensive set of results for $\Upsilon$ available to date for transitions in He~II are those of  Kisielius et al. \cite{kbn96}, the
limitations of which we have already stated. Therefore, in this paper we revisit  He~II  to improve the data available for electron collisional excitation.
Additionally, we report radiative rates (A-values) for all allowed and intercombination transitions, i.e. electric dipole (E1), electric quadrupole (E2), magnetic dipole (M1), and magnetic
quadrupole (M2),  required for complete and reliable  plasma models.

\section{Methods of calculations}

For the generation of  wavefunctions (i.e. to determine the atomic structure) we adopt the fully relativistic {\sc grasp}
(general-purpose relativistic atomic structure package) code of Grant et al. \cite{grasp0}, which has been significantly updated by Dr. P. H. Norrington, and is hosted at the website: \\ {\tt http://amdpp.phys.strath.ac.uk/UK\_APAP/codes.html}. For our calculations, the $n \leq$ 5 configurations have been considered which give rise to 25 fine-structure levels,  listed in Table 1. Furthermore,  the option of {\em extended average level} (EAL), in which a weighted (proportional to 2$j$+1) trace of the Hamiltonian matrix is minimized, is used.  This
produces a compromise set of orbitals describing closely-lying states with  moderate accuracy, although other options such as average level (AL) yield comparable results. One of the advantages of adopting this version of the code  is that its output is compatible with the input of the  Dirac atomic $R$-matrix code ({\sc darc}), used for the subsequent calculation of  collision strengths ($\Omega$), and in turn the effective collision strengths $\Upsilon$. This code has been written by  P.H.~Norrington and  I.P.~Grant, is unpublished, but is available at the same website as for {\sc grasp}. It includes relativistic effects in a systematic way, in both the target description and the scattering model, and is based on the $jj$ coupling scheme, with the Dirac-Coulomb Hamiltonian in the $R$-matrix approach. However, because of the inclusion of fine-structure in the definition of channel coupling, the matrix size of the Hamiltonian increases substantially with increasing number of levels. We also note that He~II is a light ion for which relativistic effects are not too important, but a very small degeneracy among its levels necessitates the use of a relativistic code so that subsequent results for $\Omega$ can be obtained for {\em all} possible transitions.

For our primary calculations to determine atomic parameters, the two codes discussed above have been adopted, which are  sufficient to yield accurate results. However, to assess their accuracy additional calculations have  been performed with the  {\em Flexible Atomic Code} ({\sc fac}) of Gu \cite{fac}, available from the website {\tt https://www-amdis.iaea.org/FAC/}. This is also a fully relativistic code which provides a variety of atomic parameters, and yields data which in some instances are  comparable to those generated with {\sc grasp} and {\sc {\sc darc}}. For the scattering calculations, the code is based on the {\em distorted-wave} (DW) method, which is more suitable for comparatively highly-charged ions, as  confirmed by the good agreement for a majority of transitions between the two calculations for a number of Fe ions (see Aggarwal et al. \cite{fe26b} and references therein), and another H-like ion, namely Ar~XVIII (Aggarwal et al. \cite{ar18}). Some advantages of this code are its easy adoptability and high efficiency. Furthermore, the code allows various options (such as the predefined energy grid) for calculating $\Omega$, but we have preferred to adopt the default options in which the energy grid is automatically chosen,  for two reasons. Firstly, our experience with a number of ions shows that this is a more reliable choice, and secondly and more importantly, calculations with this code are only for the purpose of comparison and accuracy assessment, because our primary  results are from  {\sc grasp} and {\sc darc}. Although, for the determination of $\Upsilon$, resonances through the {\em isolated resonance approximation} can be obtained, major aim with FAC is  to  calculate  background values of $\Omega$, so that accuracy assessments can be made. This is desirable because of the paucity of similar data for a majority of transitions in He~II.

One of the difficulties in using {\sc darc} or  {\sc fac}  is in the determination of $\Omega$ for transitions among the degenerate levels of a configuration -- see Table~1. This is because  for He~II the degeneracy is very small (i.e. $\Delta$E $\sim$ 0) and therefore the values of $\Omega$ (being dependent on $\Delta$E$_{ij}$) are highly sensitive to $\Delta$E for the `elastic' (allowed) transitions, i.e. those with $\Delta{n}$ = 0. Although both {\sc darc} and {\sc fac}  include the contribution of higher neglected partial waves from the Coulomb-Bethe formulation of Burgess et al. \cite{bht},  differences in values of $\Omega$ can sometimes be appreciable. Therefore, to resolve the discrepancies between the {\sc fac}
and {\sc darc} calculations, and to determine values of $\Omega$ as accurately as possible, we have performed yet another calculation using a combination of the close-coupling
(CC) and Coulomb-Born (CB) programs, described fully by  Igarashi et al. \cite{sn1, sn2}, and Hamada et al. ~\cite{ham}. These calculations are similar to those for elastic transitions in Al~XIII  \cite{al13c}, Ar~XVIII \cite{ar18} and Fe~XXVI  \cite{fe26b}. Furthermore, for our CC+CB calculations we have adopted the energy levels of NIST, and particularly note that such  transitions converge very {\em slowly} with increasing number of partial waves, as previously demonstrated  by Igarashi et al. {\cite{sn1} -- see also fig.~2 of \cite{ham}.  

\section{Energy levels} 

Our calculated energies  from the {\sc grasp} code, {\em with} and {\em without}  including QED effects, are given in Table~1 along with those from the experimental compilation of NIST ({\tt http://physics.nist.gov/PhysRefData}). The inclusion of QED effects {\em slightly} lowers the energies (by $\sim$ 0.000~03 Ryd), but brings these closer to the experimental results. In the case of Coulomb energies, levels with the same $n$ and angular momentum $J$ (such as 2/3 and 5/6) are quasi-degenerate, but split with the inclusion of QED effects (Lamb shift). As a result, the level orderings have changed slightly. However, we have retained the original orderings of the Coulomb energies, as these are the ones adopted in the subsequent tables. In general, the theoretical energies agree very well with the experimental values, both in magnitude and orderings.
Also listed in this table are the energies obtained with FAC. Although differences with those from GRASP are insignificant, the FAC energies are slightly closer to those of NIST.

\begin{table}
\caption{Energy levels (in Ryd) of He~II.}
\small 
\centering
\begin{tabular}{rllrrrrrrr} \hline        
Index  & Configuration & Level  &  NIST      &  GRASP$^a$  &  GRASP$^b$   & FAC$^c$ \\
\hline                                    
 1 & 1s  & $^2$S$_{1/2}$        &   0.000~000 &  0.000~000  &  0.000~000   &  0.000~000   \\
 2 & 2s  & $^2$S$_{1/2}$        &   2.999~707 &  3.000~146  &  3.000~122   &  2.999~839   \\
 3 & 2p  & $^2$P$^o$$_{1/2}$    &   2.999~702 &  3.000~146  &  3.000~118   &  2.999~831   \\  
 4 & 2p  & $^2$P$^o$$_{3/2}$    &   2.999~756 &  3.000~200  &  3.000~172   &  2.999~884   \\
 5 & 3p  & $^2$P$^o$$_{1/2}$    &   3.555~225 &  3.555~745  &  3.555~717   &  3.555~382   \\
 6 & 3s  & $^2$S$_{1/2}$        &   3.555~226 &  3.555~745  &  3.555~718   &  3.555~385   \\
 7 & 3d  & $^2$D$_{3/2}$        &   3.555~241 &  3.555~761  &  3.555~733   &  3.555~398   \\
 8 & 3p  & $^2$P$^o$$_{3/2}$    &   3.555~241 &  3.555~761  &  3.555~733   &  3.555~398   \\
 9 & 3d  & $^2$D$_{5/2}$        &   3.555~246 &  3.555~766  &  3.555~738   &  3.555~403   \\
10 & 4p  & $^2$P$^o$$_{1/2}$    &   3.749~656 &  3.750~202  &  3.750~174   &  3.749~823   \\
11 & 4s  & $^2$S $_{1/2}$       &   3.749~656 &  3.750~202  &  3.750~175   &  3.749~824   \\
12 & 4d  & $^2$D$_{3/2}$        &   3.749~662 &  3.750~209  &  3.750~181   &  3.749~830   \\
13 & 4p  & $^2$P$^o$$_{3/2}$    &   3.749~662 &  3.750~209  &  3.750~181   &  3.749~830   \\
14 & 4d  & $^2$D$_{5/2}$        &   3.749~664 &  3.750~211  &  3.750~183   &  3.749~832   \\
15 & 4f  & $^2$F$^o$$_{5/2}$    &   3.749~664 &  3.750~211  &  3.750~183   &  3.749~832   \\
16 & 4f  & $^2$F$^o$$_{7/2}$    &   3.749~665 &  3.750~212  &  3.750~184   &  3.749~833   \\
17 & 5p  & $^2$P$^o$$_{1/2}$    &   3.839~648 &  3.840~207  &  3.840~179   &  3.839~821   \\
18 & 5s  & $^2$S$_{1/2}$        &   3.839~648 &  3.840~207  &  3.840~179   &  3.839~821   \\
19 & 5d  & $^2$D$_{3/2}$        &   3.839~652 &  3.840~211  &  3.840~183   &  3.839~824   \\
20 & 5p  & $^2$P$^o$$_{3/2}$    &   3.839~652 &  3.840~211  &  3.840~183   &  3.839~824   \\
21 & 5f  & $^2$F$^o$$_{5/2}$    &   3.839~653 &  3.840~212  &  3.840~184   &  3.839~825   \\
22 & 5d  & $^2$D$_{5/2}$        &   3.839~653 &  3.840~212  &  3.840~184   &  3.839~825   \\
23 & 5g  & $^2$G$_{7/2}$        &   3.839~653 &  3.840~212  &  3.840~184   &  3.839~826   \\
24 & 5f  & $^2$F$^o$$_{7/2}$    &   3.839~653 &  3.840~212  &  3.840~184   &  3.839~826   \\
25 & 5g  & $^2$G$_{9/2}$        &   3.839~654 &  3.840~213  &  3.840~185   &  3.839~826   \\
\hline  
\end{tabular}
\begin{flushleft}
{\small
NIST: {\tt http://physics.nist.gov/PhysRefData} \\
$a$: Coulomb energies obtained with the {\sc grasp} code \\
$b$: QED corrected energies obtained with the {\sc grasp} code \\
$c$: Energies calculated with the {\sc fac} code \\
}
\end{flushleft}
\end{table}

\section{Radiative rates}

The absorption oscillator strength ($f_{ij}$) and radiative rate A$_{ji}$ (in s$^{-1}$) for a transition $i \to j$ are related by the following expression:

\begin{equation}
f_{ij} = \frac{mc}{8{\pi}^2{e^2}}{\lambda^2_{ji}} \frac{{\omega}_j}{{\omega}_i}A_{ji}
 = 1.49 \times 10^{-16} \lambda^2_{ji} (\omega_j/\omega_i) A_{ji}
\end{equation}
where $m$ and $e$ are the electron mass and charge, respectively, $c$ is the velocity of light, 
$\lambda_{ji}$ is the transition wavelength in \AA, and $\omega_i$ and $\omega_j$ are the statistical weights of the lower $i$ and upper $j$ levels, respectively.
Similarly, the oscillator strength $f_{ij}$ (dimensionless) and the line strength $S$ (in atomic unit, 1 a.u. = 6.460$\times$10$^{-36}$ cm$^2$ esu$^2$) are related by the 
 standard equations below.

\begin{flushleft}
For the electric dipole (E1) transitions 
\end{flushleft} 
\begin{equation}
A_{ji} = \frac{2.0261\times{10^{18}}}{{{\omega}_j}\lambda^3_{ji}} S^{E1} \hspace*{0.5 cm} {\rm and} \hspace*{0.5 cm} 
f_{ij} = \frac{303.75}{\lambda_{ji}\omega_i} S^{E1}, \\
\end{equation}
\begin{flushleft}
for the magnetic dipole (M1) transitions  
\end{flushleft}
\begin{equation}
A_{ji} = \frac{2.6974\times{10^{13}}}{{{\omega}_j}\lambda^3_{ji}} S^{M1} \hspace*{0.5 cm} {\rm and} \hspace*{0.5 cm}
f_{ij} = \frac{4.044\times{10^{-3}}}{\lambda_{ji}\omega_i} S^{M1}, \\
\end{equation}
\begin{flushleft}
for the electric quadrupole (E2) transitions
\end{flushleft}
\begin{equation}
A_{ji} = \frac{1.1199\times{10^{18}}}{{{\omega}_j}\lambda^5_{ji}} S^{E2} \hspace*{0.5 cm} {\rm and} \hspace*{0.5 cm}
f_{ij} = \frac{167.89}{\lambda^3_{ji}\omega_i} S^{E2}, 
\end{equation}

\begin{flushleft}
and for the magnetic quadrupole (M2) transitions 
\end{flushleft}
\begin{equation}
A_{ji} = \frac{1.4910\times{10^{13}}}{{{\omega}_j}\lambda^5_{ji}} S^{M2} \hspace*{0.5 cm} {\rm and} \hspace*{0.5 cm}
f_{ij} = \frac{2.236\times{10^{-3}}}{\lambda^3_{ji}\omega_i} S^{M2}. \\
\end{equation}

\setcounter{table}{1} 
\begin{table}
\caption{Transition wavelengths ($\lambda_{ij}$ in \AA), radiative rates (A$_{ji}$ in s$^{-1}$), oscillator strengths (f$_{ij}$, dimensionless), and line     
strengths (S, in atomic units) for electric dipole (E1), and A$_{ji}$ for E2, M1 and M2 transitions in He~II. The last column gives the ratio R  of A(E1) from our GRASP calculations and those available in the CHIANTI database. ($a{\pm}b \equiv a{\times}$10$^{{\pm}b}$).}
\small 
\centering
\begin{tabular}{rrrrrrrrrr} \hline        
$i$ & $j$ & $\lambda_{ij}$ & A$^{{\rm E1}}_{ji}$  & f$^{{\rm E1}}_{ij}$ & S$^{{\rm E1}}$ & A$^{{\rm E2}}_{ji}$  & A$^{{\rm M1}}_{ij}$ & A$^{{\rm M2}}$ & R  \\
\hline                                    
    1 &    2 &  3.037$+$02 &  0.000$+$00 &  0.000$+$00 &  0.000$+$00 &  0.000$+$00 &  2.556$-$03 &  0.000$+$00 & ----- \\
    1 &    3 &  3.037$+$02 &  1.003$+$10 &  1.387$-$01 &  2.774$-$01 &  0.000$+$00 &  0.000$+$00 &  0.000$+$00 & 1.001 \\
    1 &    4 &  3.037$+$02 &  1.003$+$10 &  2.774$-$01 &  5.548$-$01 &  0.000$+$00 &  0.000$+$00 &  1.200$+$01 & 1.000 \\
    1 &    5 &  2.563$+$02 &  2.677$+$09 &  2.636$-$02 &  4.449$-$02 &  0.000$+$00 &  0.000$+$00 &  0.000$+$00 & 1.001 \\
    1 &    6 &  2.563$+$02 &  0.000$+$00 &  0.000$+$00 &  0.000$+$00 &  0.000$+$00 &  1.136$-$03 &  0.000$+$00 & ----- \\
    1 &    7 &  2.563$+$02 &  0.000$+$00 &  0.000$+$00 &  0.000$+$00 &  3.802$+$04 &  7.102$-$06 &  0.000$+$00 & ----- \\
    1 &    8 &  2.563$+$02 &  2.677$+$09 &  5.273$-$02 &  8.897$-$02 &  0.000$+$00 &  0.000$+$00 &  4.499$+$00 & 0.999 \\
    1 &    9 &  2.563$+$02 &  0.000$+$00 &  0.000$+$00 &  0.000$+$00 &  3.802$+$04 &  0.000$+$00 &  0.000$+$00 & ----- \\
    1 &   10 &  2.430$+$02 &  1.092$+$09 &  9.662$-$03 &  1.546$-$02 &  0.000$+$00 &  0.000$+$00 &  0.000$+$00 & 1.002 \\
........ \\
........ \\
........ \\
\hline  
\end{tabular}
\end{table}

In Table 2 we present transition wavelengths ($\lambda_{ij}$ in \AA), radiative rates (A$_{ji}$ in s$^{-1}$), oscillator strengths ($f_{ij}$, dimensionless), and line strengths
($S$ in a.u.), in length form only, for all 90 electric dipole (E1) transitions among the 25 levels of He~II. The indices used to represent the lower and upper levels of a
transition have already been defined in Table 1. However, for the 107 electric quadrupole (E2), 86  magnetic dipole (M1), and 94 magnetic quadrupole (M2) transitions, only the
A-values are listed. Corresponding results for the f- and S- values may be obtained by using the above equations. 

The only other results available in the literature for comparison purposes are those  on the {\sc chianti} database for (some of) the E1 transitions. These A-values have
been determined from the calculations of Parpia and  Johnson {\cite{pj}, and there are no discrepancies with the present results for any of the transitions in common. This is evident from the last column of Table~2, where  the ratio (R) of our results and those in {\sc chianti}  are listed.  Similarly, A-values are also available from the calculations of Pal'chikov \cite{vgp}, but only for 13 E1  transitions, belonging to the $n \le$ 4 levels, for which there are no disagreements. Since these comparisons are rather limited,  we have performed another calculation from  {\sc fac}. As for the energy levels, there are no discrepancies between the two sets of A-values from  {\sc fac} and {\sc grasp}  for any of the transitions  in Table~2. Therefore, we may confidently state that the A-values listed in Table 2 are accurate to $\sim$1\%.

\section{Collision strengths}

The $R$-matrix radius has been set to 44.0 au,  and 56
continuum orbitals have been included for each channel angular momentum for the expansion of the wavefunction. This allows us to compute $\Omega$ up to an energy of 9 Ryd. The
maximum number of channels for a partial wave is 110, and the corresponding size of the Hamiltonian matrix is 6198. To obtain convergence of $\Omega$ for all
transitions and at all energies, we have included all partial waves with angular momentum $J \le$ 60, although a larger range would have been preferable for the convergence of
allowed transitions, in particular those with $\Delta n$ = 0. However, to account for higher neglected partial waves, we have included a top-up, based on the Coulomb-Bethe 
approximation for allowed  transitions and geometric series for forbidden ones.

\begin{figure}
\centering
\includegraphics[angle=-90,width=0.90\textwidth]{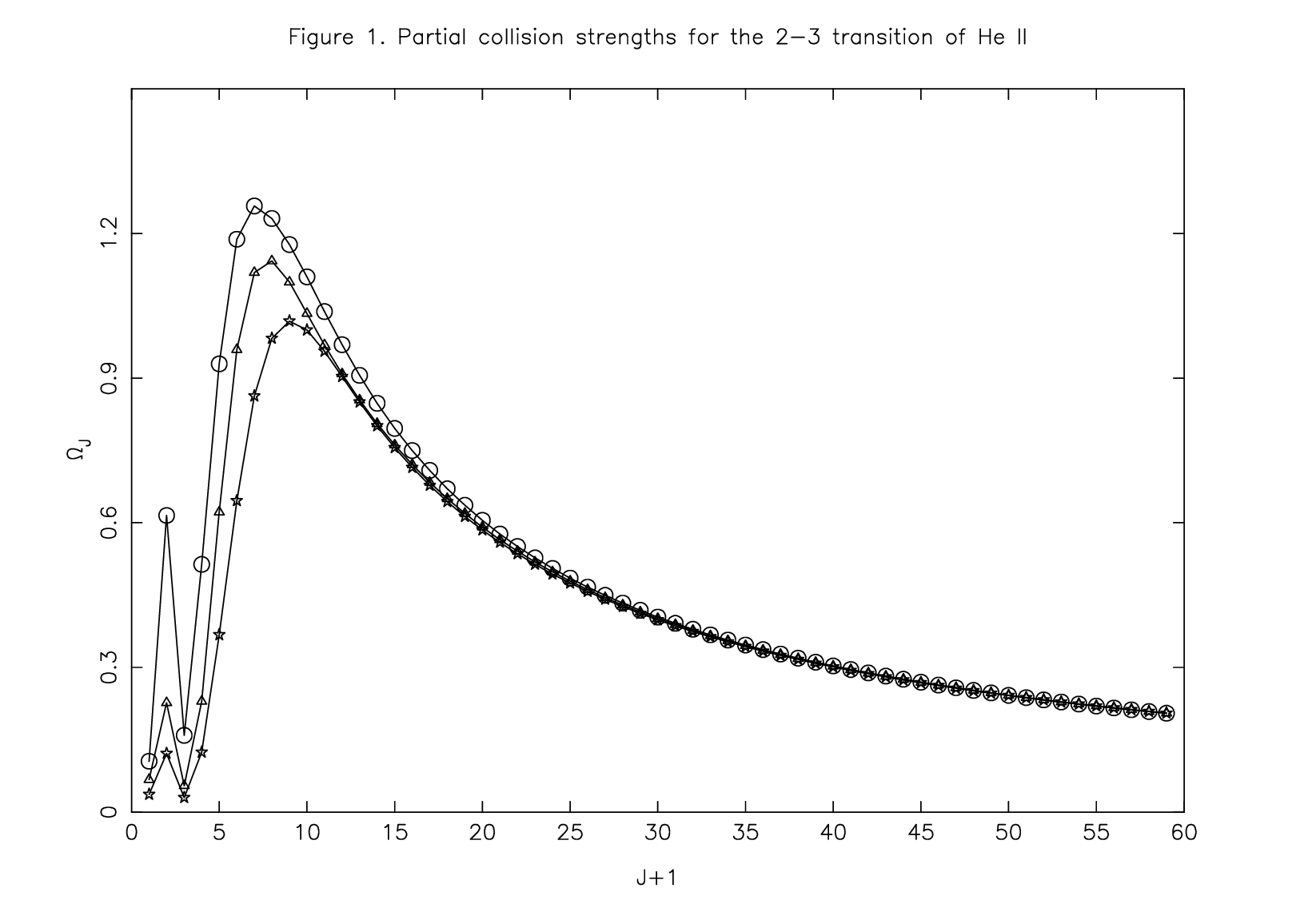}
\vspace{-0.0 cm}
\caption{Partial collision strengths for the 2s $^2$S$_{1/2}$ - 2p $^2$P$^o_{1/2}$ (2-3) transition of He~II, 
at three energies of: 4 Ryd (circles), 6 Ryd (triangles) and 8 Ryd (stars).}
\end{figure}   

\vspace{-0.0 cm}
\setcounter{figure} {1}
\begin{figure}
\centering
\includegraphics[angle=-90,width=0.90\textwidth]{fig2}
\vspace{-0.0 cm}
\caption{Partial collision strengths for the 2s $^2$S$_{1/2}$ - 3p $^2$P$^o_{1/2}$ (2-5) transition of He~II, 
at three energies of: 4 Ryd (circles), 6 Ryd (triangles) and 8 Ryd (stars).}
\end{figure}   

\vspace{-0.0 cm}
\setcounter{figure} {2}
\begin{figure}
\centering
\includegraphics[angle=-0,width=0.90\textwidth]{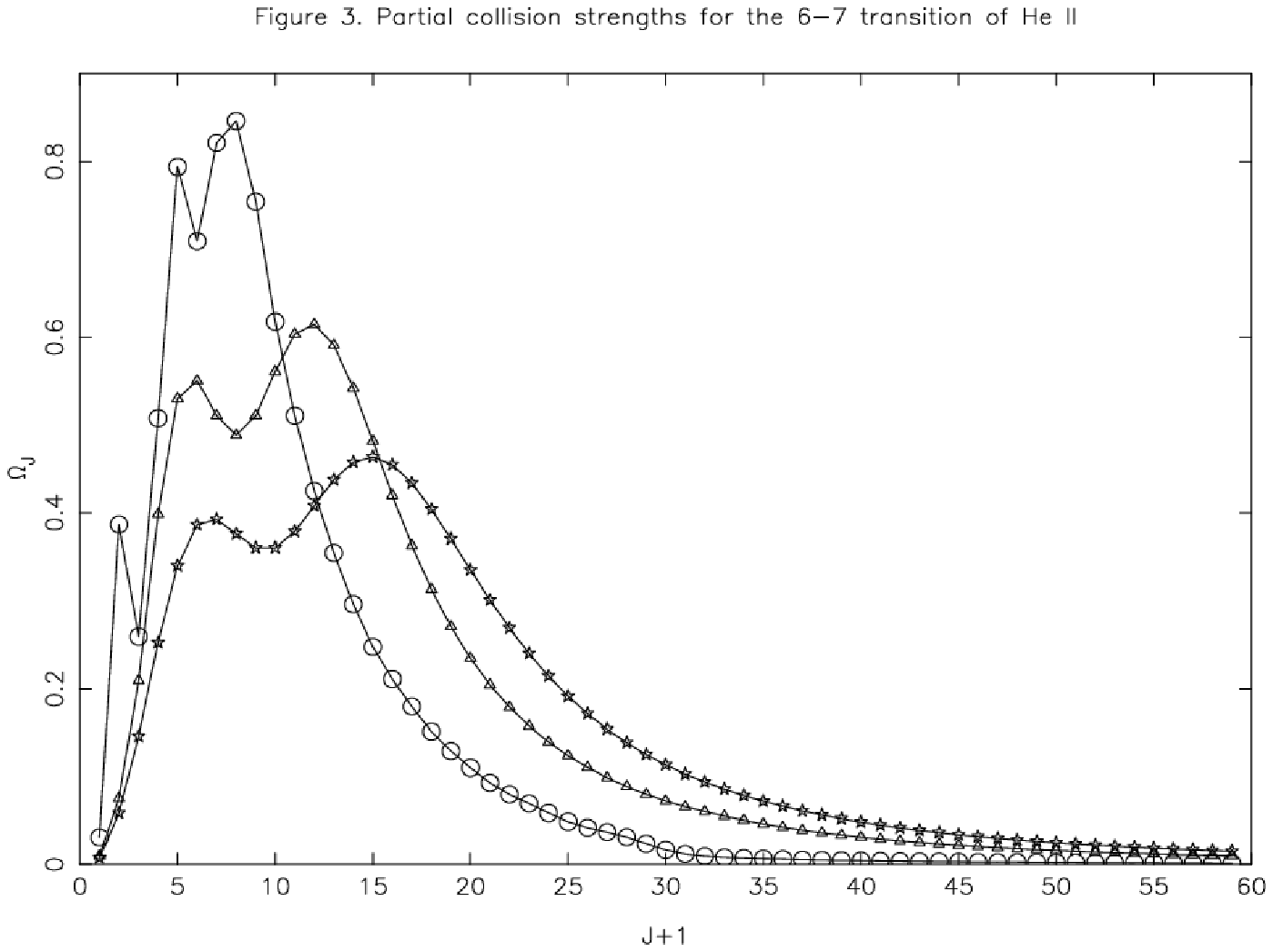}
\vspace{-0.0 cm}
\caption{Partial collision strengths for the 3s $^2$S$_{1/2}$ - 3d $^2$D$_{3/2}$ (6-7) transition of He~II, 
at three energies of:  4 Ryd (circles), 6 Ryd (triangles) and 8 Ryd (stars).}
\end{figure} 

In Figs.~1-3 we show the variation of $\Omega$ with angular momentum $J$ at three energies of 4, 6 and 8 Ryd, and for three transitions, namely 2-3 (2s $^2$S$_{1/2}$ - 2p
$^2$P$^o_{1/2}$), 2-5 (2s $^2$S$_{1/2}$ - 3p $^2$P$^o_{1/2}$) and 6-7 (3s $^2$S$_{1/2}$ - 3d $^2$D$_{3/2}$), which are `elastic' (i.e. allowed with  $\Delta n$ = 0), allowed
($\Delta n \neq$ 0), and forbidden, respectively. For the forbidden and allowed transitions shown in Figs.~2 and 3, the values of $\Omega$ have  fully converged at all energies,
including the highest energy of our calculations.  However, for the `elastic' transitions our range of partial waves is not sufficient for the convergence of $\Omega$, as shown
in Fig. 1. For such transitions the top-up from  the Coulomb-Bethe approximation is quite significant.

In Table~3 we present our results of $\Omega$ for all transitions over a wider energy range (4 $\le$ E $\le$ 9 Ryd), but above thresholds. The indices adopted to represent  a 
transition are given in Table~1. These results for $\Omega$ are not directly applicable in plasma modelling, but are very useful in assessing the  accuracy of a
calculation. Unfortunately there are no other similar results for the fine-structure transitions, although values of $\Omega$ are available for the $LS$ transitions in the
energy range 4 $\le$ E $\le$ 7 Ryd from our earlier calculations  \cite{he2a}. Therefore, we have performed another calculation using the {\sc fac} code 
 to make an accuracy assessment for the present results.

The values of $\Omega$ calculated from the {\sc fac} code are listed in Table~3 at a single {\em excited} (E$_j$) energy of $\sim$ 5.5 Ryd, which nearly corresponds to the
highest (initial) energy of our calculations, i.e. 9 Ryd. These $\Omega$  from {\sc fac} provide a ready comparison with our corresponding results from {\sc darc}. 
The two sets of $\Omega$ from {\sc darc} and {\sc fac} generally agree within 20\% for a majority of transitions. However, for about 20\% of the transitions (examples
include 2-23, 3-23 and 4-23) there are differences of over 20\%. In particular, six transitions, namely 1 (1s $^2$S$_{1/2}$) - 15 (4f $^2$F$^o_{5/2}$), 16 (4f $^2$F$^o_{7/2}$),
21 (5f $^2$F$^o_{5/2}$),  23 (5g $^2$G$_{7/2}$), 24 (5f $^2$F$^o_{7/2}$), and 25 (5g $^2$G$_{9/2}$), have larger discrepancies of up to (almost) an order of magnitude.  All of
these are resonance transitions but their values of $\Omega$ are very small, i.e. $\le$ 10$^{-4}$, and hence may not affect  plasma modelling applications. Such differences, for weak transitions, between the two codes are common and have been noted for a wide range of ions. They partly arise due to the different methodologies adopted (i.e. $R$-matrix and DW) and partly because the FAC code does not calculate $\Omega$ at each partial wave, but rather interpolates at regular intervals  to improve efficiency.  Nevertheless, for
comparatively stronger transitions there is good agreement between the two independent calculations, as also illustrated in Fig.~4 for the 5-7 (3p $^2$P$^o_{1/2}$ - 3d
$^2$D$_{3/2}$), 6-8 (3s $^2$S$_{1/2}$ - 3p $^2$P$^o_{3/2}$) and 8-9 (3p $^2$P$^o_{3/2}$ - 3d $^2$D$_{5/2}$) transitions.

\setcounter{table}{2} 
\begin{table}
\caption{Collision strengths for transitions in  He~II from our calculations with the DARC code. Presently calculated results from FAC are also listed at a single incident energy of $\sim$9~Ryd. ($a{\pm}b \equiv$ $a\times$10$^{{\pm}b}$).}
\small 
\centering
\begin{tabular}{rrllllllllll} \hline        
\multicolumn{2}{c}{Transition} & \multicolumn{7}{c}{Energy (Ryd)}\\                                           
\hline                                                                                                        
  $i$ & $j$ & 4 &   5 &   6 &   7 &   8 &   9 & FAC \\ 
\hline       
  1 &  2 &  1.200$-$01 &  1.423$-$01 &  1.486$-$01 &  1.551$-$01 &  1.606$-$01 &  1.667$-$01 &  1.979$-$01 \\
  1 &  3 &  1.690$-$01 &  2.619$-$01 &  3.183$-$01 &  3.619$-$01 &  4.020$-$01 &  4.388$-$01 &  4.742$-$01 \\
  1 &  4 &  3.380$-$01 &  5.237$-$01 &  6.365$-$01 &  7.238$-$01 &  8.039$-$01 &  8.774$-$01 &  9.482$-$01 \\
  1 &  5 &  3.047$-$02 &  5.816$-$02 &  6.672$-$02 &  7.281$-$02 &  7.844$-$02 &  8.382$-$02 &  8.403$-$02 \\
  1 &  6 &  3.828$-$02 &  3.896$-$02 &  3.395$-$02 &  3.252$-$02 &  3.210$-$02 &  3.249$-$02 &  3.920$-$02 \\
  1 &  7 &  1.289$-$02 &  1.282$-$02 &  1.088$-$02 &  1.017$-$02 &  9.971$-$03 &  9.989$-$03 &  7.994$-$03 \\
  1 &  8 &  6.090$-$02 &  1.163$-$01 &  1.334$-$01 &  1.456$-$01 &  1.568$-$01 &  1.676$-$01 &  1.681$-$01 \\
  1 &  9 &  1.934$-$02 &  1.922$-$02 &  1.632$-$02 &  1.526$-$02 &  1.496$-$02 &  1.498$-$02 &  1.199$-$02 \\
  1 & 10 &  1.425$-$02 &  2.551$-$02 &  2.760$-$02 &  2.913$-$02 &  3.069$-$02 &  3.226$-$02 &  3.016$-$02 \\
........ \\
........ \\
........ \\
\hline  
\end{tabular}
\end{table}

\vspace{-0.0 cm}
\setcounter{figure} {3}
\begin{figure}
\centering
\includegraphics[angle=-90,width=0.90\textwidth]{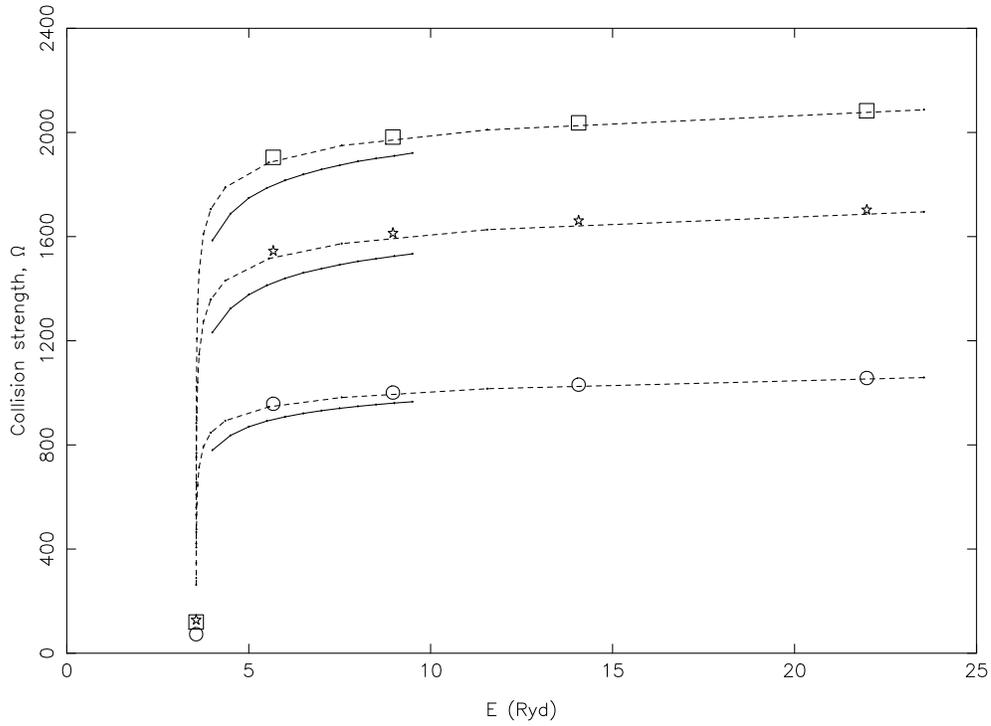}
\vspace{-0.0 cm}
\caption{Comparison of collision strengths for the 5-7 (3p $^2$P$^o_{1/2}$ - 3d $^2$D$_{3/2}$), 6-8 (3s $^2$S$_{1/2}$ - 3p $^2$P$^o_{3/2}$) and 8-9 
(3p $^2$P$^o_{3/2}$ - 3d $^2$D$_{5/2}$) transitions of He~II. Continuous curves are from our calculations from {\sc darc}, broken curves are from the CC+CB programs, 
and the $\Omega$ from FAC are shown as, circles: 5-7, stars: 6-8 and squares: 8-9 transition.}
\end{figure}

However, there are four allowed transitions, namely 14-15 (4d $^2$D$_{5/2}$ - 4f $^2$F$^o_{5/2}$), 19-20 (5d $^2$D$_{3/2}$ - 5p $^2$P$^o_{3/2}$),  21-22 (5f $^2$F$^o_{5/2}$ - 5d
$^2$D$_{5/2}$), and 23-24 (5g $^2$G$_{7/2}$ - 5f $^2$F$^o_{7/2}$), for which the discrepancies between the {\sc fac} and {\sc darc} calculations are comparatively more significant. The energy differences ($\Delta$E) for these transitions are (almost) zero in our {\sc grasp} and {\sc fac} calculations as shown in Table~1, and hence the correspondingly values of $\Omega$ are rather sensitive as stated in section~2. Therefore,  to resolve the differences between the {\sc fac} and {\sc darc} calculations, and to determine values of $\Omega$ as accurately as possible, we have performed yet another calculation using the CC+CB code and the energy levels of NIST.

The results for $\Omega$ obtained from  CC+CB  are shown in Fig.~4 for the 5-7, 6-8 and 8-9 transitions, and agree very well with the other two calculations from {\sc
darc} and {\sc fac}. This provides  confidence in the calculated values of $\Omega$ from  CC+CB. In Fig. 5 we show similar comparisons between $\Omega$ values from
 {\sc fac} and CC+CB  for three transitions, namely 14-15 (4d $^2$D$_{5/2}$ - 4f $^2$F$^o_{5/2}$), 21-22 (5f $^2$F$^o_{5/2}$ - 5d $^2$D$_{5/2}$) and 23-24 (5g
$^2$G$_{7/2}$ - 5f $^2$F$^o_{7/2}$). For these transitions  there are also no (great) discrepancies between the {\sc fac} and CC+CB calculations, although the FAC results are lower by $\sim$10\% for 14--15. Therefore, for the 26 elastic transitions we have adopted values of $\Omega$ from our CC+CB calculations, and from {\sc darc}  for the  remaining 274.

\setcounter{figure} {4}
\begin{figure}
\centering
\includegraphics[angle=-90,width=0.90\textwidth]{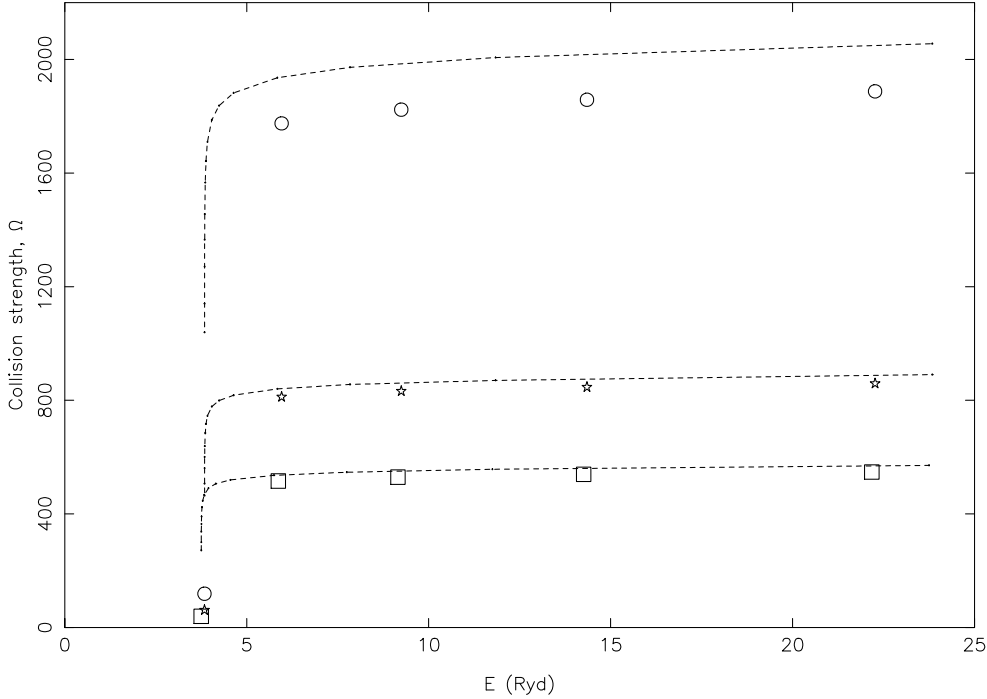}
\vspace{-0.0 cm}
\caption{Comparison of collision strengths for the 14-15 (4d $^2$D$_{5/2}$ - 4f $^2$F$^o_{5/2}$),  21-22 (5f $^2$F$^o_{5/2}$ - 5d $^2$D$_{5/2}$) and 23-24 (5g $^2$G$_{7/2}$ -
5f $^2$F$^o_{7/2}$) transitions of He~II. Broken curves are from the CC+CB programs, and the $\Omega$ from FAC are shown as, squares: 14-15, circles: 21-22 and stars: 
23-24 transition.}
\end{figure}

\section{Effective collision strengths}

Effective collision strengths $\Upsilon$ are obtained after integrating $\Omega$ over a Maxwellian distribution of electron velocities, i.e.

\begin{equation}
\Upsilon(T_e) = \int_{0}^{\infty} {\Omega}(E) \,exp(-E_j/kT_e) d(E_j/{kT_e})
\end{equation}
where E$_j$ is the incident energy of the electron with respect to the final state of the transition, $k$ is Boltzmann's constant, and T$_e$ is the electron temperature in K.
 Once the value of $\Upsilon$ is known for a transition, the corresponding value of the excitation $q(i,j)$ and de-excitation $q(j,i)$ rate coefficients can be easily
  obtained from the following simple relations:

\begin{equation}
q(i,j) = \frac{8.63 \times 10^{-6}}{{\omega_i}{T_e^{1/2}}} \Upsilon \,{\rm exp}(-E_{ij}/{kT_e}) \hspace*{1.0 cm}{\rm cm^3s^{-1}}
\end{equation}
and

\begin{equation}
q(j,i) = \frac{8.63 \times 10^{-6}}{{\omega_j}{T_e^{1/2}}} \Upsilon \hspace*{1.0 cm}{\rm cm^3 s^{-1}},
\end{equation}
where $\omega_i$ and $\omega_j$ are the statistical weights of the initial ($i$) and final ($j$) states, respectively, and E$_{ij}$ is the transition energy.

\setcounter{figure} {5}
\begin{figure*}
\includegraphics[angle=90,width=0.90\textwidth]{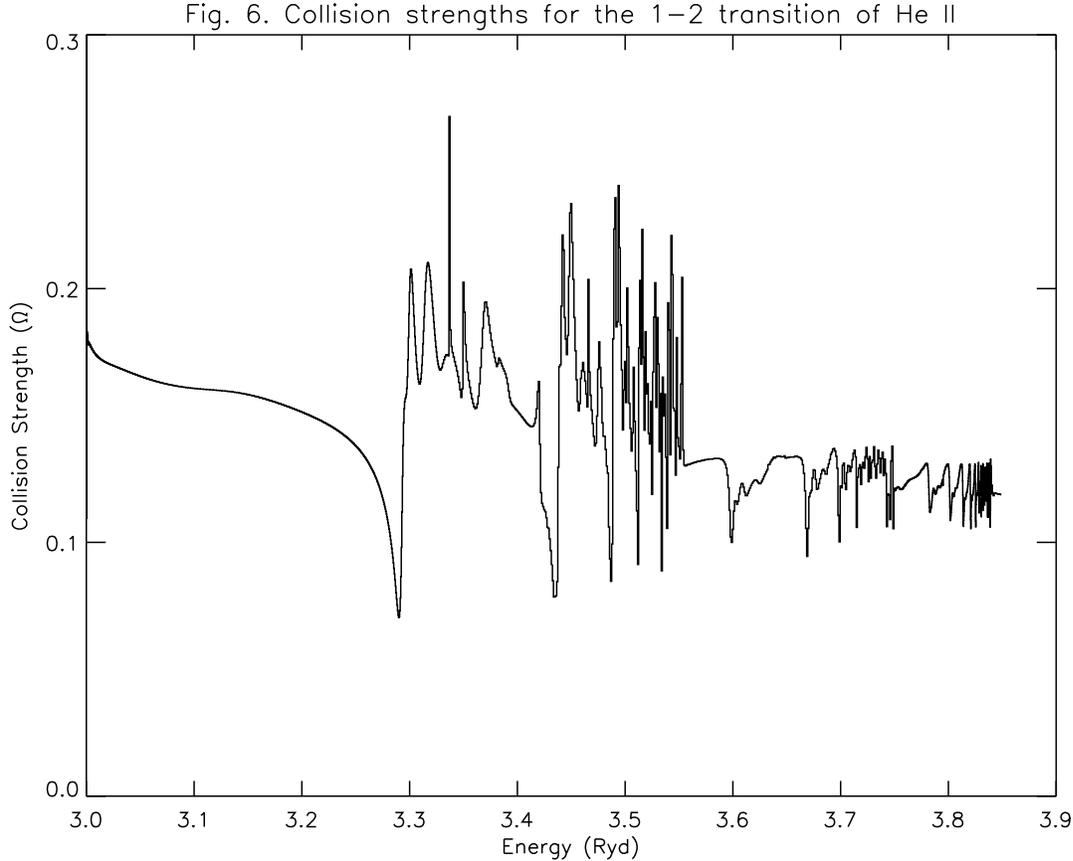}
\vspace{-0.0 cm}
\caption{Collision strengths for the 1s $^2$S$_{1/2}$ - 2s $^2$S$_{1/2}$ (1-2) transition of He~II.}
\end{figure*}

\setcounter{figure} {6}
\begin{figure*}
\includegraphics[angle=90,width=0.90\textwidth]{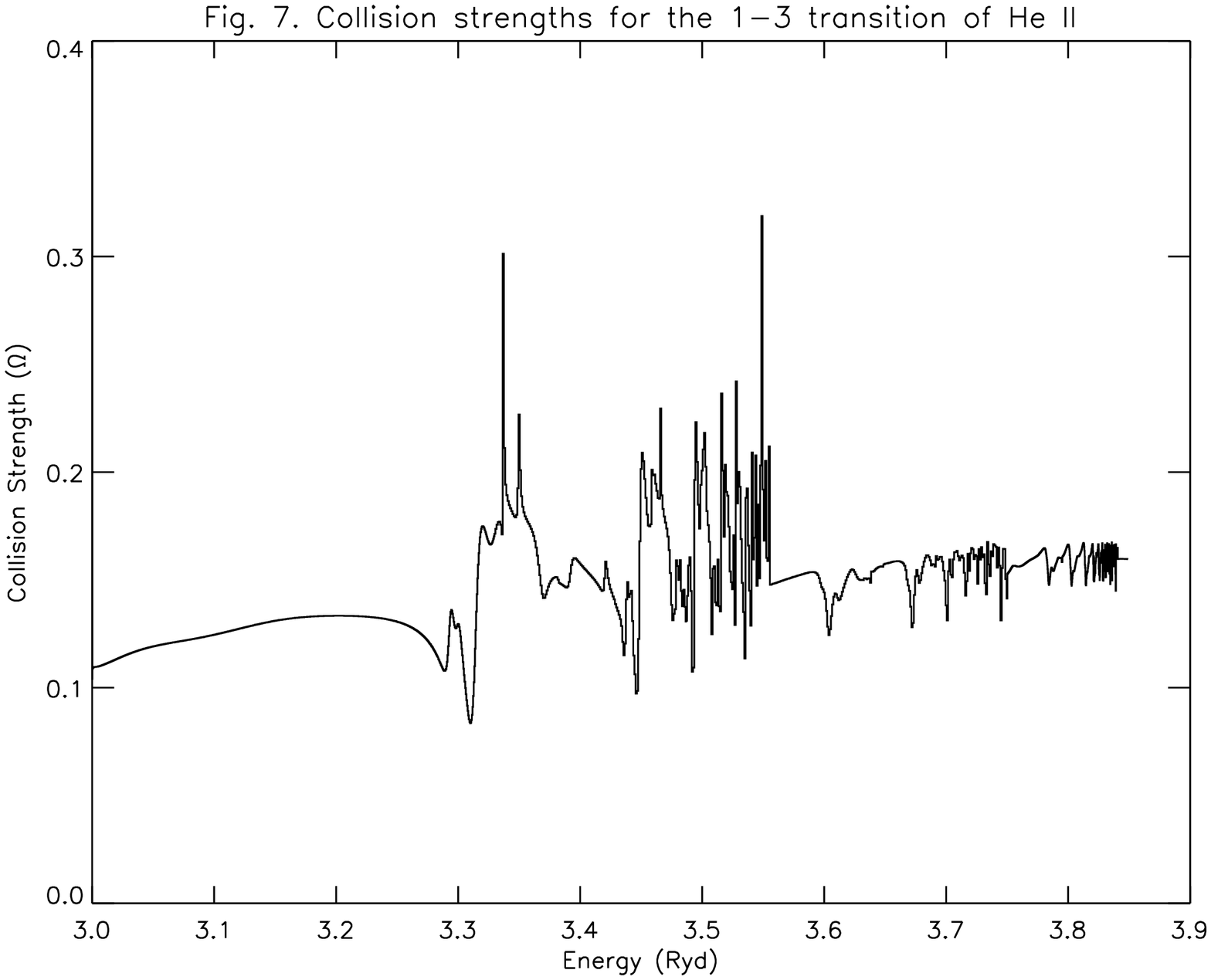}
\vspace{-0.0 cm}
\caption{Collision strengths for the 1s $^2$S$_{1/2}$ - 2p $^2$P$^o_{1/2}$ (1-3) transition of He~II.}
\end{figure*}

\setcounter{figure} {7}
\begin{figure*}
\includegraphics[angle=90,width=0.90\textwidth]{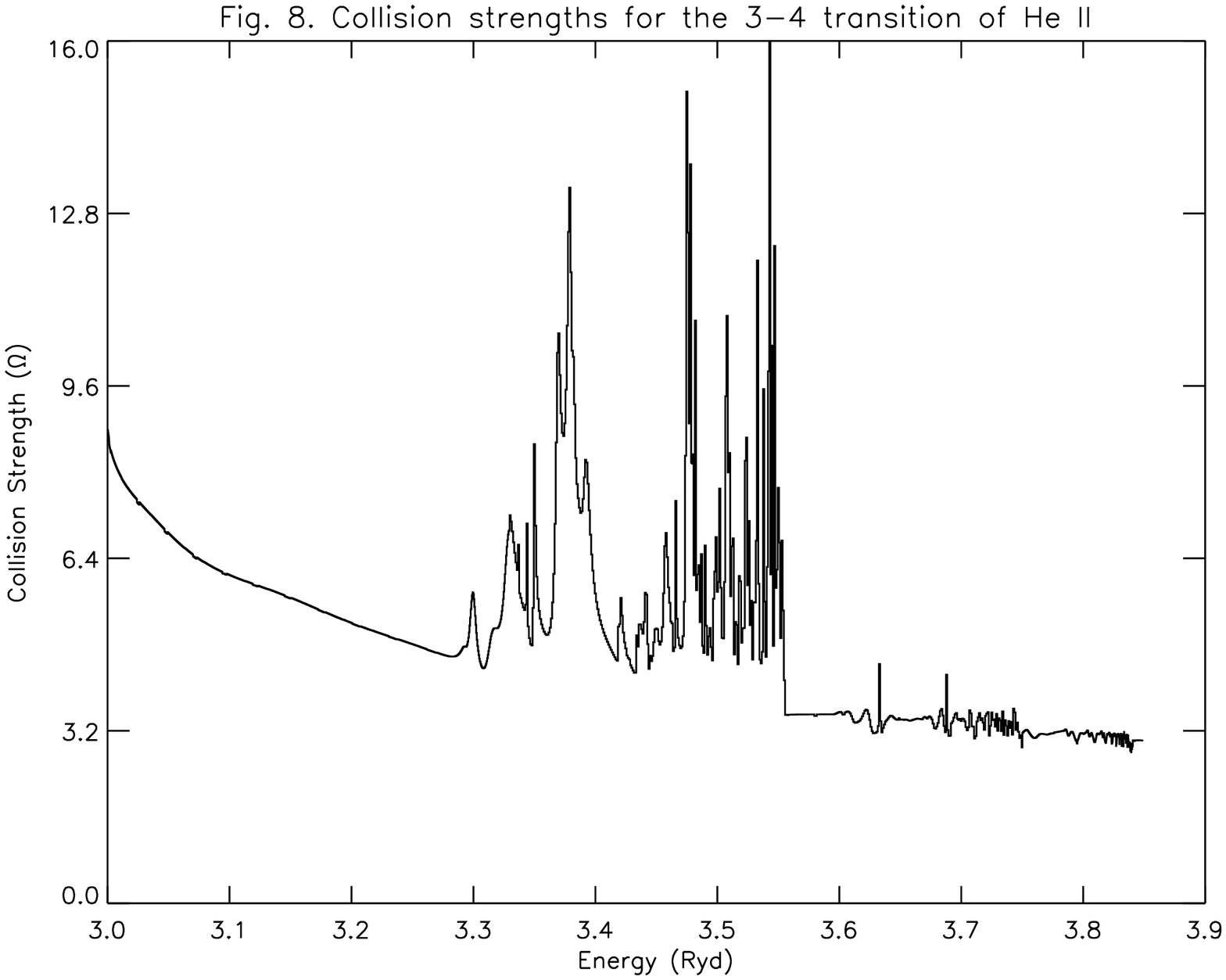}
\vspace{-0.0 cm}
\caption{Collision strengths for the 2p $^2$P$^o_{1/2}$ - 2p $^2$P$^o_{3/2}$ (3-4) transition of He~II.}
\end{figure*}

Since the threshold energy region is dominated by numerous resonances, values of $\Omega$  have been computed at a large number of energies  to delineate these
resonances. We have performed our calculations of $\Omega$ at $\sim$ 3000 energies in the threshold region. However, the energy mesh is not uniform and varies between 0.000~01
and 0.001 Ryd. The density and importance of resonances can be appreciated from Figs.~6-8, where we show our $\Omega$ values in the thresholds region for the 1-2 (1s
$^2$S$_{1/2}$ - 2s $^2$S$_{1/2}$), 1-3 (1s $^2$S$_{1/2}$ - 2p $^2$P$^o_{1/2}$), and 3-4 (2p $^2$P$^o_{1/2}$ - 2p $^2$P$^o_{3/2}$) transitions, respectively. Resonance structure
for the 1-2 transition is similar to that shown  in Fig.~1 of our earlier calculations in $LS$ coupling  \cite{he2a},  and there are only a few minor differences  in the magnitudes and positions of resonances, which do not  significantly affect the $\Upsilon$ results. However,  the discrepancies with the
measurements of Dolder and  Peart \cite{dp}  remain the same, i.e. the experimental results are underestimated in comparison to theory by $\sim$ 30\%.

Our calculated values of $\Upsilon$ are listed in Table~4 over a wide temperature range of 3.6 $\le$ log T$_e \le$ 5.4 K, suitable for applications to astrophysical and other
plasmas. As noted in section~1, the most comprehensive similar calculations available in the literature for comparison  are those of Kisielius et al.
\cite{kbn96}. For a majority of transitions there is good agreement between the two calculations over the entire range of common temperatures, i.e. 3.2 $\le$ log T$_e$ (K)
$\le$ 4.3. However, six transitions (4-15, 7-14, 10-14, 10-15, 11-14, and 11-15) show  discrepancies of over 20\%. We illustrate this in Fig.~9 for three transitions, namely 7-14
(3d $^2$D$_{3/2}$ - 4d $^2$D$_{5/2}$), 10-15 (4p $^2$P$^o_{1/2}$ - 4f $^2$F$^o_{5/2}$) and 11-15 (4s $^2$S$_{1/2}$ - 4f $^2$F$^o_{5/2}$). All  six transitions are {\em
forbidden}, but converge slowly as demonstrated in Fig.~10 for the 11-15 transition, for which the discrepancy is the greatest. As a result of slow convergence, $\Omega$ values
for these transitions either decrease slowly with increasing energy or remain nearly constant, as shown in Table~3 and also verified by the FAC calculations. Since
the discrepancies between our values of $\Upsilon$ and those listed by  Kisielius et al. increase with increasing temperature, it appears that their calculations of $\Omega$
at higher energies are comparatively less accurate. However, as Kisielius et al. have not reported any data for collision strengths, we cannot directly verify the differences between
the corresponding values of $\Upsilon$. Furthermore, an exercise performed by including only $J \le$ 40  (i.e. the partial waves range adopted by these authors), shows a difference of
less than 20\% between the $\Omega$ values so obtained and those in Table~3. Therefore, the reason for the differences between the two sets of $\Upsilon$ values may
lie somewhere else. We discuss this further below.

Apart from the  six transitions listed above, there are three others, namely 12-14 (4d $^2$D$_{3/2}$ - 4d $^2$D$_{5/2}$), 13-15 (4p $^2$P$^o_{3/2}$ - 4f $^2$F$^o_{5/2}$) and
15-16 (4f $^2$F$^o_{5/2}$ - 4f $^2$F$^o_{7/2}$), for which the discrepancy between our results of $\Upsilon$ and those of Kisielius et al. \cite{kbn96} is up to two orders of
magnitude over the entire range of common temperatures, with the $\Upsilon$ values of Kisielius et al. invariably higher. For these transitions, there are no discrepancies
between our calculations from {\sc darc} and {\sc fac}, and our values of $\Upsilon$ (especially at higher temperatures) closely follow the corresponding results for $\Omega$.
Therefore, we are confident that our  results for $\Upsilon$ are comparatively more accurate than those of Kisielius et al.

\evensidemargin -1.2cm 
\oddsidemargin -1.2cm

\setcounter{table}{3} 
\begin{table}
\caption{Effective collision strengths for transitions in  He~II. ($a{\pm}b \equiv a{\times}10^{{\pm}b}$).}
\small 
\centering
\begin{tabular}{rrllllllllll} \hline        
\multicolumn {2}{c}{Transition} & \multicolumn{10}{c}{Temperature (log T$_e$, K)}\\                                          
\hline                                                                                                        
$i$ & $j$ & 3.60 &   3.80 &   4.00 &   4.20  &  4.40 &   4.60 &   4.80 &   5.00 &   5.20 &   5.40  \\
\hline       
  1 &  2 &  1.713$-$01 &  1.689$-$01 &  1.661$-$01 &  1.629$-$01 &  1.595$-$01 &  1.553$-$01 &  1.502$-$01 &  1.458$-$01 &  1.432$-$01 &  1.417$-$01 \\
  1 &  3 &  1.141$-$01 &  1.162$-$01 &  1.191$-$01 &  1.229$-$01 &  1.281$-$01 &  1.349$-$01 &  1.445$-$01 &  1.597$-$01 &  1.831$-$01 &  2.118$-$01 \\
  1 &  4 &  2.264$-$01 &  2.311$-$01 &  2.371$-$01 &  2.449$-$01 &  2.555$-$01 &  2.692$-$01 &  2.885$-$01 &  3.190$-$01 &  3.660$-$01 &  4.233$-$01 \\
  1 &  5 &  2.232$-$02 &  2.306$-$02 &  2.398$-$02 &  2.482$-$02 &  2.589$-$02 &  2.795$-$02 &  3.157$-$02 &  3.679$-$02 &  4.309$-$02 &  4.897$-$02 \\
  1 &  6 &  4.615$-$02 &  4.616$-$02 &  4.619$-$02 &  4.558$-$02 &  4.439$-$02 &  4.309$-$02 &  4.198$-$02 &  4.095$-$02 &  3.970$-$02 &  3.777$-$02 \\
  1 &  7 &  1.967$-$02 &  1.905$-$02 &  1.853$-$02 &  1.785$-$02 &  1.695$-$02 &  1.601$-$02 &  1.518$-$02 &  1.444$-$02 &  1.369$-$02 &  1.277$-$02 \\
  1 &  8 &  4.415$-$02 &  4.574$-$02 &  4.767$-$02 &  4.942$-$02 &  5.161$-$02 &  5.576$-$02 &  6.304$-$02 &  7.350$-$02 &  8.613$-$02 &  9.789$-$02 \\
  1 &  9 &  2.964$-$02 &  2.867$-$02 &  2.786$-$02 &  2.680$-$02 &  2.543$-$02 &  2.402$-$02 &  2.277$-$02 &  2.166$-$02 &  2.053$-$02 &  1.915$-$02 \\
  1 & 10 &  1.626$-$02 &  1.492$-$02 &  1.399$-$02 &  1.362$-$02 &  1.394$-$02 &  1.498$-$02 &  1.664$-$02 &  1.867$-$02 &  2.082$-$02 &  2.248$-$02 \\
........ \\
........ \\
........ \\
\hline  
\end{tabular}
\end{table}

\setcounter{figure} {8}
\begin{figure*}
\includegraphics[angle=-90,width=0.90\textwidth]{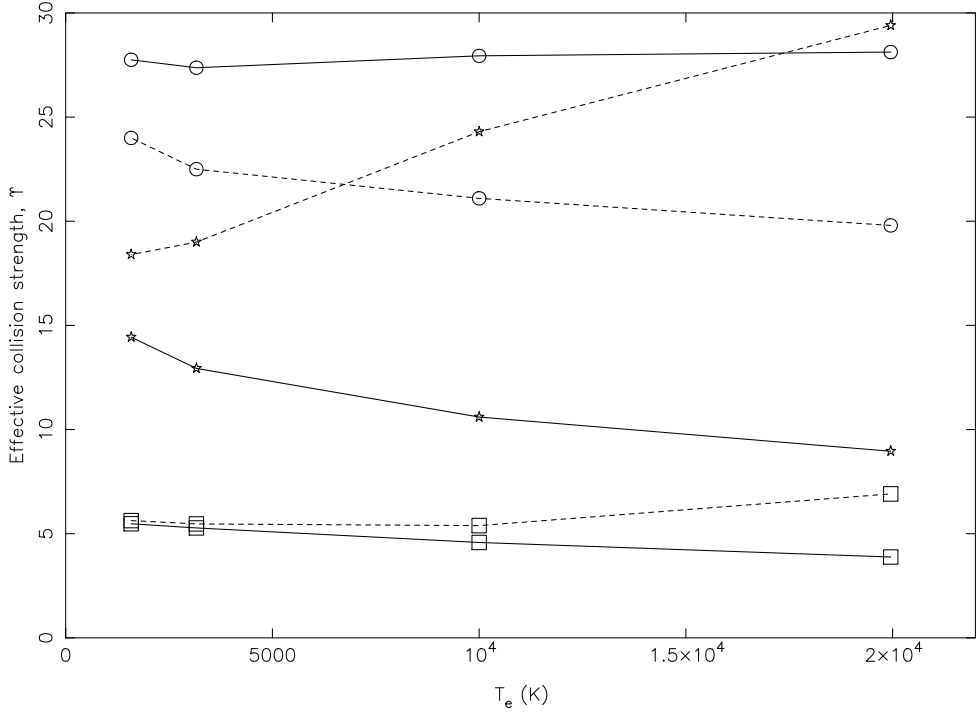}
\vspace{-0.0 cm}
\caption{Comparison of effective collision strengths for the 7-14 (squares: 3d $^2$D$_{3/2}$ - 4d $^2$D$_{5/2}$), 10-15 (circles: 4p $^2$P$^o_{1/2}$ - 4f $^2$F$^o_{5/2}$) and
11-15 (stars: 4s $^2$S$_{1/2}$ - 4f $^2$F$^o_{5/2}$) transitions of He~II. Continuous curves are the present results from {\sc darc} and broken curves are from 
Kisielius et  al.  \cite{kbn96}.}
\end{figure*}

\setcounter{figure} {9}
\begin{figure*}
\includegraphics[angle=-90,width=0.90\textwidth]{fig10}
\vspace{-0.0 cm}
\caption{Partial collision strengths for the 4s $^2$S$_{1/2}$ - 4f $^2$F$^o_{5/2}$ (11-15) transition of He~II, 
at three energies of:  4 Ryd (circles), 6 Ryd (triangles) and 8 Ryd (stars).}
\end{figure*}

\setcounter{figure} {10}
\begin{figure*}
\includegraphics[angle=90,width=0.90\textwidth]{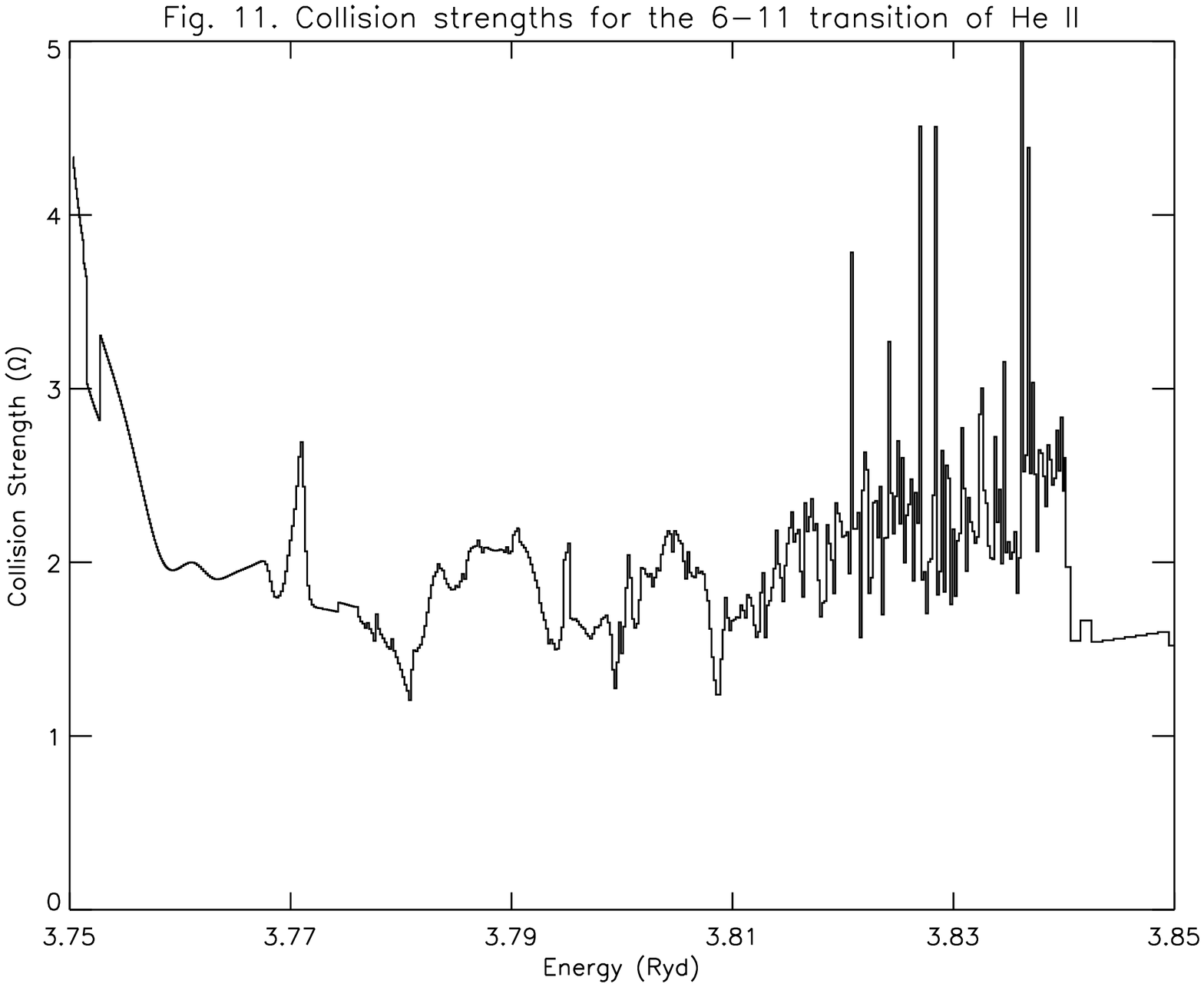}
\vspace{-0.0 cm}
\caption{Collision strengths for the 3s $^2$S$_{1/2}$ - 4s $^2$S$_{1/2}$ (6-11) transition of He~II.}
\end{figure*}

Kisielius et al. \cite{kbn96} compared their values of $\Upsilon$ for some  $LS$ transitions with our earlier calculations  \cite{he2a}, and observed
differences of up to a factor of three. For almost all the transitions  in their Table~2, their values of $\Upsilon$ are not only higher but are also in agreement with the
present calculations, hence confirming the inaccuracy of our earlier results in $LS$ coupling. However, they explained the differences between the two calculations on the basis
of the resonances they observed near the thresholds, which we do not believe to be the correct explanation.  In Table~5a we compare 
values of $\Omega$ from both our present and previous calculations  \cite{he2a} for some transitions for which the discrepancies are large. Corresponding values
of $\Upsilon$ are listed in Table~5b, but only for our current work  and that of  Kisielius et al. Earlier results from Aggarwal et al.   \cite{he2a} are listed at a single temperature of
10$^4$~K to provide a ready comparison. For the 1s-4s and 1s-4p transitions, the two sets of $\Omega$ agree at energies above 4~Ryd, but at E = 4~Ryd the earlier values of
$\Omega$ are considerably lower. We have no explanation for this as the earlier data are no longer available. However,  it does explain the smaller values of $\Upsilon$ obtained in the earlier calculations, as it affects the results more at lower temperatures than the higher ones. For the 1s-4d, 3s-4s, and 3s-4p transitions, the two sets of $\Omega$  agree over the
entire common energy range of 4 $\le$ E $\le$ 7~Ryd, yet the earlier results for $\Upsilon$ are lower by up to a factor of two. We  observe resonances for these (corresponding
fine-structure) transitions, as shown in Fig.~11 for  6-11 (3s $^2$S$_{1/2}$ - 4s $^2$S$_{1/2}$). However, these (and similar other) resonances do not have a
significant effect on  $\Upsilon$, because the energy range of the  near threshold resonances is very narrow ($<$ 0.1~Ryd). As a result, 
$\Upsilon$  closely follow the background values of $\Omega$ over the entire temperature range. Therefore, the differences between the $LS$ and $jj$ values of $\Upsilon$
are not due to resonances, although we are unable to fully understand these. Finally, for the 4s-4f and 4p-4f transitions, not only do our values of $\Omega$ from the earlier
$LS$ calculations agree with the present work, but the corresponding  $\Upsilon$ also agree. Clearly, for these two (and some other) transitions, the $\Upsilon$ values
of Kisielius et al. are not only anomalous but must also be  in error.

\setcounter{table}{4}                                                                                                                                           
\begin{table*}                                                                                                                                                                                                                                                                                                   
\caption{a. Comparison of collision strengths ($\Omega$) for some transitions of  He~II.}    
\begin{tabular}{lllllllll}                                                                                                                                      
\hline                                                                                                                                                                                                                                                                                                             
Energy     & \multicolumn{2}{c}{4.0} & \multicolumn{2}{c}{5.0} & \multicolumn{2}{c}{6.0} & \multicolumn{2}{c}{7.0 Ryd}  \\                                                         
            \cline{2-9}   
Transition &  DARC & RM              &  DARC & RM              &  DARC & RM              &   DARC & RM                  \\
\hline                                                                                                                                  
1s - 4s &  0.0242 & 0.0157 & 0.0212 & 0.0235 & 0.0157 & 0.0159 & 0.0137 & 0.0130 \\
1s - 4p &  0.0427 & 0.0315 & 0.0765 & 0.0740 & 0.0828 & 0.0785 & 0.0874 & 0.0863 \\
1s - 4d &  0.0257 & 0.0268 & 0.0194 & 0.0199 & 0.0148 & 0.0144 & 0.0133 & 0.0133 \\
3s - 4s &  3.258  & 3.220  & 8.902  & 9.199  & 10.52  & 10.61  & 11.13  & 11.09  \\
3s - 4p &  4.336  & 4.252  & 13.28  & 13.84  & 19.38  & 22.64  & 23.97  & 29.34  \\
4s - 4f &  13.08  & 16.80  & 15.51  & 16.79  & 14.69  & 14.99  & 14.17  & 14.12  \\
4p - 4f &  93.03  & 107.0  & 115.6  & 116.0  & 113.1  & 108.0  & 109.3  & 100.7  \\
\hline                                                                                                                                      
\end{tabular}                                                                                                                               
\end{table*}                                                                                      

\setcounter{table}{4}
\begin{table*}
 \caption{b. Comparison of effective collision strengths ($\Upsilon$) for some transitions of  He~II. ($a{\pm}b \equiv a{\times}10^{{\pm}b}$).}                                 
\begin{tabular}{llllllllll}                                                                                                               
\hline                                                                                                                                                                                                                                                                          
Transition & \multicolumn{4}{c}{DARC} & \multicolumn{4}{c}{DRM} & RM  \\                                                         
            \cline{2-10} 
T$_e$ (log K)  &  3.2 & 3.5 & 4.0 & 4.3 & 3.2 & 3.5 & 4.0 & 4.3 &  4.0 \\
\hline 
1s - 4s &  3.51-2 & 3.03-2 & 2.57-2 & 2.46-2 &  3.63-2 & 3.11-2 & 2.56-2 & 2.44-2 &  8.63-3 \\ 
1s - 4p &  6.04-2 & 5.11-2 & 4.19-2 & 4.10-2 &  6.25-2 & 5.24-2 & 4.25-2 & 4.16-2 &  1.55-2 \\
1s - 4d &  4.41-2 & 3.75-2 & 3.06-2 & 2.84-2 &  4.54-2 & 3.84-2 & 3.06-2 & 2.80-2 &  1.56-2 \\
3s - 4s &  2.51-0 & 2.24-0 & 2.11-0 & 2.43-0 &  2.59-0 & 2.28-0 & 2.14-0 & 2.49-0 &  1.37-0 \\
3s - 4p &  7.04-0 & 6.47-0 & 5.48-0 & 5.08-0 &  7.19-0 & 6.57-0 & 5.98-0 & 6.14-0 &  2.90-0 \\
4s - 4f &  3.49+1 & 3.13+1 & 2.69+1 & 2.29+1 &  3.84+1 & 3.65+1 & 3.88+1 & 4.20+1 &  2.48+1 \\
4p - 4f &  121.1  & 118.0  & 124.5  & 119.1  &  1588   & 1607   &  1624  & 1602   &  119.9  \\
\hline                                                                                                                                      
\end{tabular}                                                                                                                               
\begin{flushleft}
{\small
DARC: Present calculations \\
DRM: Calculations of Kisielius et al. \cite{kbn96} \\
RM:  Calculations of Aggarwal et al. \cite{he2a} \\
}
\end{flushleft}
\end{table*}

\setcounter{figure}{11}
\begin{figure*}
\includegraphics[angle=-90,width=0.90\textwidth]{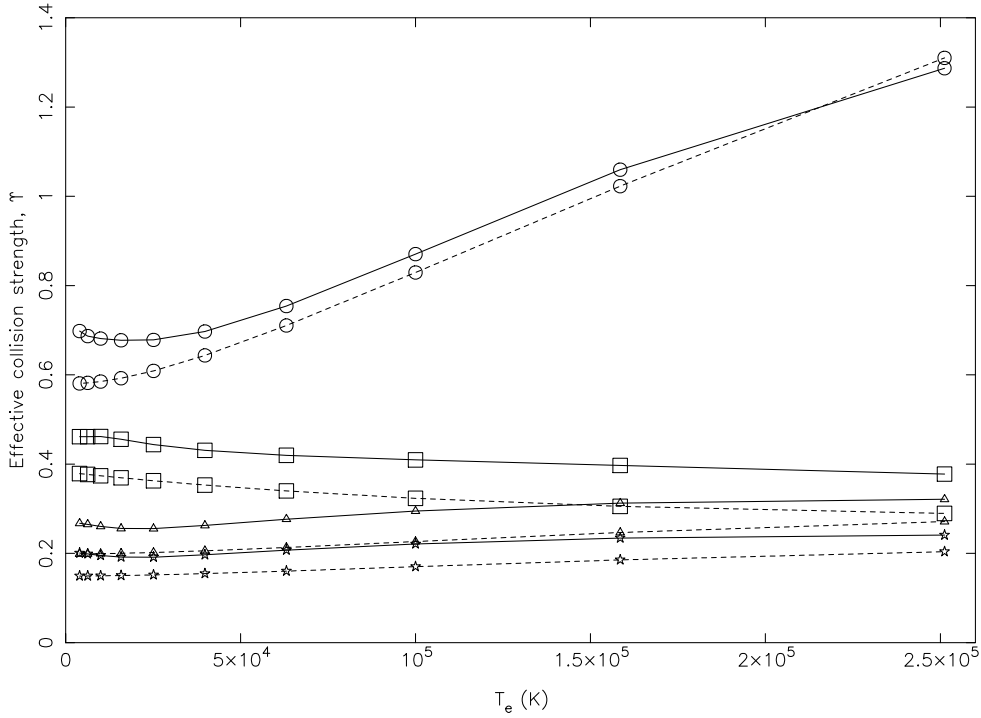}
\vspace{-0.0 cm}
\caption{Comparison of effective collision strengths for the 1-6 (squares: 1s $^2$S$_{1/2}$ - 3s $^2$S$_{1/2}$), 2-5 (circles: 2s $^2$S$_{1/2}$ - 3p $^2$P$^o_{1/2}$),
2-15 (stars: 2s $^2$S$_{1/2}$ - 4f $^2$F$^o_{5/2}$), and 2-16 (triangles: 2s $^2$S$_{1/2}$ - 4f $^2$F$^o_{7/2}$) transitions of He~II. Continuous curves are the present 
results from {\sc darc} and broken curves are from the {\sc chianti} database. Note that for clarity $\Upsilon$ for the 1--6 transition have been multiplied by a factor of 10.}
\end{figure*}

The other $\Upsilon$ values available in the literature are in the {\sc chianti} database, from the calculations of Ballance et al. \cite{bbs03}, for transitions from the lowest two levels to
higher excited levels. Unfortunately however, for a majority of the common transitions, there is no agreement between the two sets of $\Upsilon$. We demonstrate this in Fig.~12
for four transitions, namely 1-6 (1s $^2$S$_{1/2}$ - 3s $^2$S$_{1/2}$), 2-5 (2s $^2$S$_{1/2}$ - 3p $^2$P$^o_{1/2}$), 2-15 (2s $^2$S$_{1/2}$ - 4f $^2$F$^o_{5/2}$), and 2-16 (2s
$^2$S$_{1/2}$ - 4f $^2$F$^o_{7/2}$).  Since our values of $\Upsilon$ are invariably higher, by up to 30\%, it is clearly because of the inclusion of pseudostates in the expansion of the
wavefunctions by Ballance et al. This process takes account of the higher ionisation channels (neglected in the present work) which results in a loss of flux, and subsequently
to lower values of $\Omega$ (and hence $\Upsilon$), for some of the transitions, as also demonstrated  by Aggarwal et al. \cite{he2b} and  
Ballance et al.  However, differences between our results for $\Upsilon$ and those of Ballance et al. are up to a factor of four for some transitions, such as 1-19 (1s
$^2$S$_{1/2}$ - 5d $^2$D$_{3/2}$), 2-18 (2s $^2$S$_{1/2}$ - 5s $^2$S$_{1/2}$) and 2-24 (2s $^2$S$_{1/2}$ - 5f $^2$F$^o_{7/2}$). For these (and some other) transitions the
discrepancies between the two sets of $\Upsilon$ persist over the entire range of temperatures. However, in the absence of any results for $\Omega$ from the calculations of
Ballance et al., it is difficult to arrive at any definite conclusion.  

\section{Conclusions}

In the present work, results for energy levels, radiative rates, collision strengths, and effective collision strengths are listed for {\em all} transitions among the lowest 25 levels of He~II.
Radiative rates are also presented for four types of transitions, namely E1, E2, M1, and M2.
This complete dataset should be useful for modelling a variety of plasmas. 

Our calculations have been performed in the $jj$ coupling scheme, CI (configuration interaction) and relativistic effects  included while generating
wavefunctions, and a  wide range of partial waves adopted to achieve convergence in $\Omega$  for a majority of transitions. Furthermore, resonances have
been resolved in a  fine energy mesh  to improve the accuracy of the derived values of $\Upsilon$. Similarly, $\Omega$ have been computed over a wide energy range up to
9~Ryd  to determine  $\Upsilon$ up to a temperature of 10$^{5.4}$~K. 

Three other calculations, based on the same $R$-matrix method as adopted in the present work, are available in the literature. Differences among all these calculations
are significant for some of the transitions. We are able to understand and explain most of the differences, but not all,  because  data for collision
strengths are not available for the earlier calculations. Nevertheless, based on comparisons among a variety of calculations, the accuracy of our 
$\Omega$ and $\Upsilon$ is probably better than 20\% for a majority of transitions. However, scope remains for further improvement, mainly because we have not included
pseudostates in the expansion of wavefunctions. Their inclusion in a calculation may decrease  $\Upsilon$ by up to 30\%, but only for some transitions, as
demonstrated by Aggarwal et al. \cite{he2b} and Ballance et al. \cite{bbs03}. Until such calculations are performed, the present values of $\Omega$ and $\Upsilon$
are not only for {\em all} transitions among the lowest 25 levels of He~II, but are also probably the best currently available.

\section*{Acknowledgment}
We thank Dr. Kerry Lawson of CCFE, UK for bringing our attention to the importance of this work.   


\newpage

\begin{thebibliography}{999}

\bibitem{dp}     Dolder, K.T.;  Peart, B.   A measurement of cross sections for the 1S-2S excitation of He$^+$ ions by electron impact. {\em J. Phys.} {\bf 1973}, B  6, 2415.
\bibitem{he2a}   Aggarwal, K.M.; Berrington, K.A.; Kingston, A.E.; Pathak, A.  Electron collision  strengths for all transitions among the $n$=1, 2, 3, 4 and 5 levels of He$^+$.  {\em J. Phys.}  {\bf 1991}, B  24, 1757.
\bibitem{kbn96}  Kisielius, R.; Berrington, K.A.;  Norrington, P.H. Atomic data from the IRON Project. XV. Electron excitation of the fine-structure transitions in hydrogen-like ions He~II and Fe~XXVI.   {\em Astron. Astrophys. Supl.} {\bf 1996}, 118, 157.
\bibitem{bbs03}  Ballance, C.P.; Badnell, N.R.; Smyth, E.S.  A pseudo-state sensitivity study on hydrogenic ions.  {\em J. Phys.} {\bf 2003}, B 36, 3707.
\bibitem{rm1}    Berrington, K.A.; Burke, P.G.; LeDourneuf, M.; Robb, W.D.; Taylor, K.T.; Vo Ky Lan.  A new version of the general program to calculate atomic continuum processes using the R-matrix method.  {\em Comput. Phys. Commun.} {\bf 1978},  14, 367.
\bibitem{pb}     Bryans, P.; Badnell, N.R.; Gorczyca, T.W.; Laming, J.M.; Mitthumsiri, W.;  Savin, D.W. Collisional ionization equilibrium for optically thin plasmas. I. updated recombination rate coefficients for bare through sodium-like ions.    {\em Astrophys. J. Supl.} {\bf 2006}, 167, 343.
\bibitem{grasp0} Grant, I.P.; McKenzie, B.J.; Norrington, P.H.; Mayers, D.F.;  Pyper, N.C.  An atomic multiconfigurational Dirac-Fock package.  {\em Comput. Phys. Commun.} {\bf 1980},  21,  207.
\bibitem{fac}    Gu, M.F.  The flexible atomic code.  {\em Can. J. Phys.} {\bf 2008}, 86, 675.
\bibitem{fe26b} Aggarwal, K.M.; Hamada, K.; Igarashi, A.; Jonauskas, V.; Keenan, F.P.;   Nakazaki, S.  Radiative rates and electron impact excitation rates for H-like Fe~XXVI.  {\em Astron. Astrophys.} {\bf 2008}, 484, 879.
\bibitem{ar18}  Aggarwal, K.M.; Hamada, K.; Igarashi, A.; Jonauskas, V.; Keenan, F.P.;   Nakazaki, S.  Radiative rates and electron impact excitation rates for H-like Ar~XVIII.  {\em Astron. Astrophys.} {\bf 2008}, 487, 383.
\bibitem{bht}    Burgess, A.; Hummer, D.G.;  Tully, J. A. Electron impact excitation of positive ions.   {\em Phil. Trans. Roy. Soc.} {\bf 1970}, A 266, 225.
\bibitem{sn1}    Igarashi, A.; Horiguchi, Y.; Ohsaki, A.;  Nakazaki, S.  Electron-impact excitations between the n=2 fine-structure levels of hydrogenic ions.  {\em J. Phys. Soc. Japan} {\bf 2003}, 72, 307.
\bibitem{sn2}    Igarashi, A.; Ohsaki, A.;  Nakazaki, S.  Electron-exchange effect in electron-impact excitation of the n=2 fine-structure levels of hydrogenic ions.  {\em J. Phys. Soc. Japan} {\bf 2005}, 74, 321.
\bibitem{ham} Hamada, K.; Aggarwal, K.M.;  Akita, K.: Igarashi, A.;  Keenan, F.P.;   Nakazaki, S.  Effective collision strengths for optically allowed transitions among degenerate levels of hydrogenic ions with 2 $\le$ Z $\le$ 30.   {\em At. Dsata Nucl. Data Tables} {\bf 2010}, 96, 481.
\bibitem{al13c} Aggarwal, K.M.; Igarashi, A.; Keenan, F.P.;   Nakazaki, S.  Effective collision strengths for allowed transitions among the $n \le$ 5 degenerate levels of 
Al~XIII.   {\em Astron. Astrophys.} {\bf 2008}, 479, 585.
\bibitem{pj}     Parpia, F.A.;  Johnson, W.R.  Radiative decay rates of metastable one-electron atoms.  {\em Phys. Rev.} {\bf 1982}, A 26, 1142.
\bibitem{vgp}    Pal'chikov, V.G.  Relativistic transition probabilities and oscillator strengths in hydrogen-like atoms.   {\em Phys. Scr.} {\bf 1998}, 57, 581.

\bibitem{he2b}   Aggarwal, K.M.; Callaway, J.; Kingston, A.E.;   Unnikrishnan, K.  Excitation rate coefficients for transitions among the $n$ = 1, 2 and 3 levels of He$^+$.  {\em Asrtophys. J. Supl.} {\bf 1992}, 80, 473.

\end{thebibliography}
\end{document}